# Real-space titration and manipulation of particle-like correlated electrons in doped Mott insulator

Yanyan Geng[1,2,+], Haoyu Dong[1,2,+], Renhong Wang[1,2,+], Zilu Wang[1,2,5,+], Jianfeng Guo[1,2,3], Shuo Mi[1,2], Yan Li[3], Fei Pang[1,2], Rui Xu[1,2], Li Huang[3], Hong-Jun Gao[3], Wei Ji[1,2,*], Shancai Wang[1,2,*], Weichang Zhou[4,*], and Zhihai Cheng[1,2,*]

[1]*Key Laboratory of Quantum State Construction and Manipulation (Ministry of Education), School of Physics, Renmin University of China, Beijing, 100872, China*
[2]*Beijing Key Laboratory of Optoelectronic Functional Materials & Micro-nano Devices, School of Physics, Renmin University of China, Beijing 100872, China*
[3]*Beijing National Laboratory for Condensed Matter Physics, Institute of Physics, Chinese Academy of Sciences, Beijing 100190, China*
[4]*Key Laboratory of Low-dimensional Quantum Structures and Quantum Control of Ministry of Education, School of Physics and Electronics, Hunan Research Center of the Basic Discipline for Quantum Effects and Quantum Technologies, Hunan Normal University, Changsha 410081, China*
[5]*College of Science, Hainan Tropical Ocean University, Sanya 572022, China*

**Abstract:** The localized (particle-like) correlated electrons deserve particular attention as they govern various exotic quantum phenomena, such as quantum spin liquids, Wigner crystals, and Mott insulators in correlated systems. However, direct observation and manipulation of these particle-like electrons at the atomic or single-electron scale remain highly challenging. Here, we successfully realize and directly visualize particle-like correlated electrons in 1$T$-TaS$_2$ through hole doping. The potential-dependent local electronic structure of single particle-like electron is revealed by angle-resolved photoemission spectroscopy (ARPES), scanning tunneling spectroscopy (STS) combined with theoretical calculations. The complex correlated interactions including nearest-neighbor attractive interactions and many-body repulsive interactions are further demonstrated and discussed based on the spatial distribution of particle-like electrons. Furthermore, the tentative manipulation of the particle-like electrons is successfully achieved by the energy-excitation mechanism. Our results not only provide profound insights into particle-like electrons in correlated systems, but also establish a versatile platform for designing and controlling quantum states at the atomic scale.

[+]These authors contributed equally: Yanyan Geng, Haoyu Dong, Renhong Wang, Zilu Wang
* Email: wji@ruc.edu.cn, scwang@ruc.edu.cn, wchangzhou@hunnu.edu.cn, zhihaicheng@ruc.edu.cn

**Introduction**

Correlated electron systems host rich quantum states arising from the competition between itinerant (wave-like) and localized (particle-like) electrons [1-4]. In weakly correlated systems, the wave-like electrons dominate delocalized behaviors, exemplified by metals [5,6], low-dimensional Fermi liquids [7,8], and topological semimetals [9]. However, when on-site Coulomb repulsion becomes comparable to the kinetic energy scale, the system enters the strongly correlated regime [10-12]. This triggers a dynamic competition between wave-like and particle-like electrons in Fig. 1a and b, giving rise to emergent quantum states such as Wigner crystals [13,14], quantum spin liquids [15,16], and Mott insulators [17-19]. Notably, Mott insulators, characterized by the stark contradiction between partial-filled bands (wave-like electrons) and insulating ground states (particle-like electrons), serve as an ideal platform for exploring novel quantum phenomena in strongly correlated systems.

The microscopic mechanism of Mott insulators is typically described by the Hubbard model, where the competition between itinerant kinetic energy (t) and on-site Coulomb repulsion ($U$) determines the ground state of the system [20,21]. Carrier doping effectively modulates the $U$/t ratio, driving the emergence of exotic quantum phases, including high-temperature superconductivity and charge/spin/orbital ordered states [22-24]. For instance, Qiu, Z. *et al.* reported the imbalanced electron-hole crystals in a doped Mott insulator α-$RuCl_3$ [25]. Yim, C. *et al.* proposed that charge disproportionation could generate a unique insulating ground state on doped Mott insulator surfaces [26]. Similarly, in twisted graphene moiré superlattices, tuning flat-band filling factors yields various distinctive phenomena such as superconductivity [27], Wigner crystals [28-30], the fractional quantum Hall effect [31], and electron nematic phases [32].

As a typical Mott insulator, 1$T$-$TaS_2$ has garnered significant attention due to its rich and tunable electronic states [33-35]. A series of strategies, such as electrostatic gating [36], laser [37] and voltage pulses [38], surface atomic adsorption [39], doping [40], were developed to efficiently tune the electron correlation strength ($U$/t ratio) of 1$T$-$TaS_2$, making 1$T$-$TaS_2$ as an ideal platform for studying the competition between particle-like and wave-like electrons. However, these previous studies have primarily focused on the macroscopic measurement [41,42], the direct investigation of local electronic states, interactions and even manipulation of particle-like electrons in the atomic scale is still very challenging. Recently, scanning tunneling microscopy (STM) has been successfully employed for vertical and lateral manipulation of individual atoms/molecules [43], as well as the construction of artificial atomic lattices such as quantum corrals and Kagome lattices [44]. Thus, STM is considered as an ideal tool for probing and manipulating the localized particle-like electrons with atomic precision, displayed in Fig. 1c, which might open new avenues for real-space investigation and manipulation of the spatial distribution and electronic states of particle-like electrons.

**In this work,** we systematically investigate the particle-like correlated electrons in hole-doped 1$T$-TaS$_2$ using STM/STS, ARPES combined with theoretical calculations. The $\boldsymbol{k}$-space ARPES and $\boldsymbol{r}$-space STS spectra measurements reveal the doping-dependent average electronic structure and the potential-dependent localized electronic structure of single particle-like electron. The spatial organization of particle-like electrons demonstrates small electron clusters and anti-7 magic electron configurations, reflecting the nearest-neighbor attractive interactions and many-body repulsive interactions between correlated electrons. Remarkably, the tentative manipulation of the particle-like electrons and local charge density wave (CDW) ordering-disordering transition is realized by energy-excitation and electron-injection/extraction mechanisms. This work provides key results for understanding the complicated behavior and interactions of particle-like electrons and suggests new directions for achieving more exotic quantum states through electron manipulation.

# Results and discussion

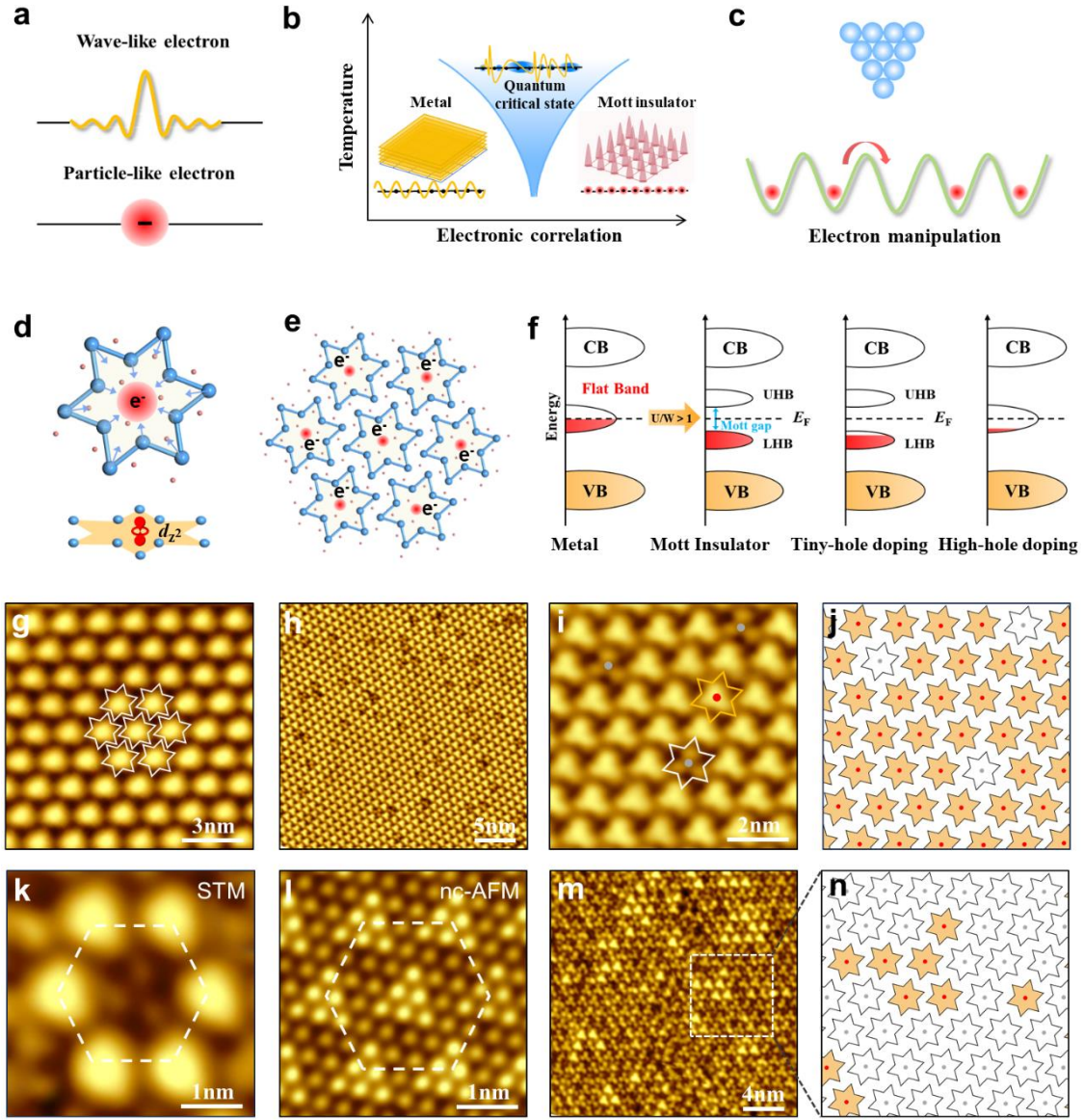

**Figure 1. Correlated electronic states of hole-doped 1$T$-TaS$_2$.** (a) Electrons exhibit particle-like and wave-like behaviors. (b) Schematic phase diagram of the metal (wave-like electron)-Mott insulator (particle-like electron) transition [45]. (c) Schematic of particle-like electron manipulation. (d) Atomic models of the Star of David (SoD) cluster with single particle-like electron localized at the central Ta atom ($d_{z^2}$ orbital). (e) Schematic of the commensurate SoD clusters in the CDW state of 1$T$-TaS$_2$. (f) Schematic band structures of Metal, Mott insulator, tiny-hole-doped and high-hole-doped Mott insulator with less particle-like electrons in the flat-band. (g) High-resolution STM image of pristine 1$T$-TaS$_2$ with the overlaid SoD models. (h-j) Large-scale (h), high-resolution (i) STM images, and their schematic model (j) of tiny-hole-doped 1$T$-TaS$_2$. The central electron-filled SoD and electron-empty star of triangular (SoT) are marked by the central red and grey dots, respectively. (k,l) Constant-current STM (k) and corresponding atomic-resolved nc-AFM (l) images of SoD and SoT. (m,n) Large-scale STM image (m) and their schematic model (n) of high-hole-doped 1$T$-TaS$_2$. Scanning parameter: (h,i,k,m) $V$=-0.6 V, $I$=-100 pA; (g) $V$=-1 V, $I$=-100 pA.

The 1$T$-TaS$_2$ exhibits a correlated Mott-insulating ground state in the commensurate CDW (CCDW) state. In the CCDW state, every 13 Ta atoms shrink into a cluster named the Star of David (SoD) as shown in Fig. 1d. The SoD reconstructs into a long coherent √13 ×√13 superlattice in the CCDW state (Fig. 1e). Each SoD contains 13 Ta 5$d$ electrons, the 12 electrons of the outer Ta atoms pair up to form six occupied CDW bands, leaving a single particle-like electron localized at the central Ta atom ($d_{z^2}$ orbital) in a half-filled metallic flat band. This half-filled band further splits into upper and lower Hubbard bands (UHB and LHB) due to the large on-site Coulomb $U$, forming a Mott insulator. Hole doping can precisely control the flat-band filling factor and effectively modulate the correlation effects in 1$T$-TaS$_2$ (Fig. 1f and Supplementary Fig. 1), providing an exemplary platform for achieving various intriguing electronic states.

At the tiny-hole-doping condition, discrete distributions of three-petal-flower stars appear within the CCDW superlattice, as illustrated in Fig. 1h-j and Supplementary Fig. 2, displaying a darker contrast compared to the normal SoDs in the occupied state. According to our previous work [34] and first-principles calculations in Supplementary Fig. 3, these emergent dark three-petal-flower stars (highlighted by the white stars with the central grey circles) are the electron-empty SoTs, contrasting with the electron-filled SoDs (highlighted by the bright stars with the central red dots). The constant-current STM and corresponding atomic-resolved nc-AFM images of SoD and SoT confirm the absence of atomic defects in SoT regions, as shown in Fig. 1k and l, further verifying the electron-empty nature of SoTs. Consequently, STM can be used as a local electron detector to directly visualize the spatial distribution of particle-like electrons (electron-filled SoDs) of hole-doped 1$T$-TaS$_2$

Figure 1m and n present the large-scale STM images and schematic model of high-hole-doped 1$T$-TaS$_2$. Only a small number of localized electron-filled SoDs remain, exhibiting particle-like electron features. Unexpectedly, these particle-like electrons aggregate into small clusters or short-chain configurations rather than distributing randomly. This particular spatial correlation implies the presence of significant attractive interactions among the particle-like electrons, rather than dominant repulsive interactions (discussed in detail below). These discrete and interactive electrons, localized within the SoDs, establish an ideal platform for the spatial investigation and manipulation of particle-like correlated electrons at the atomic scale.

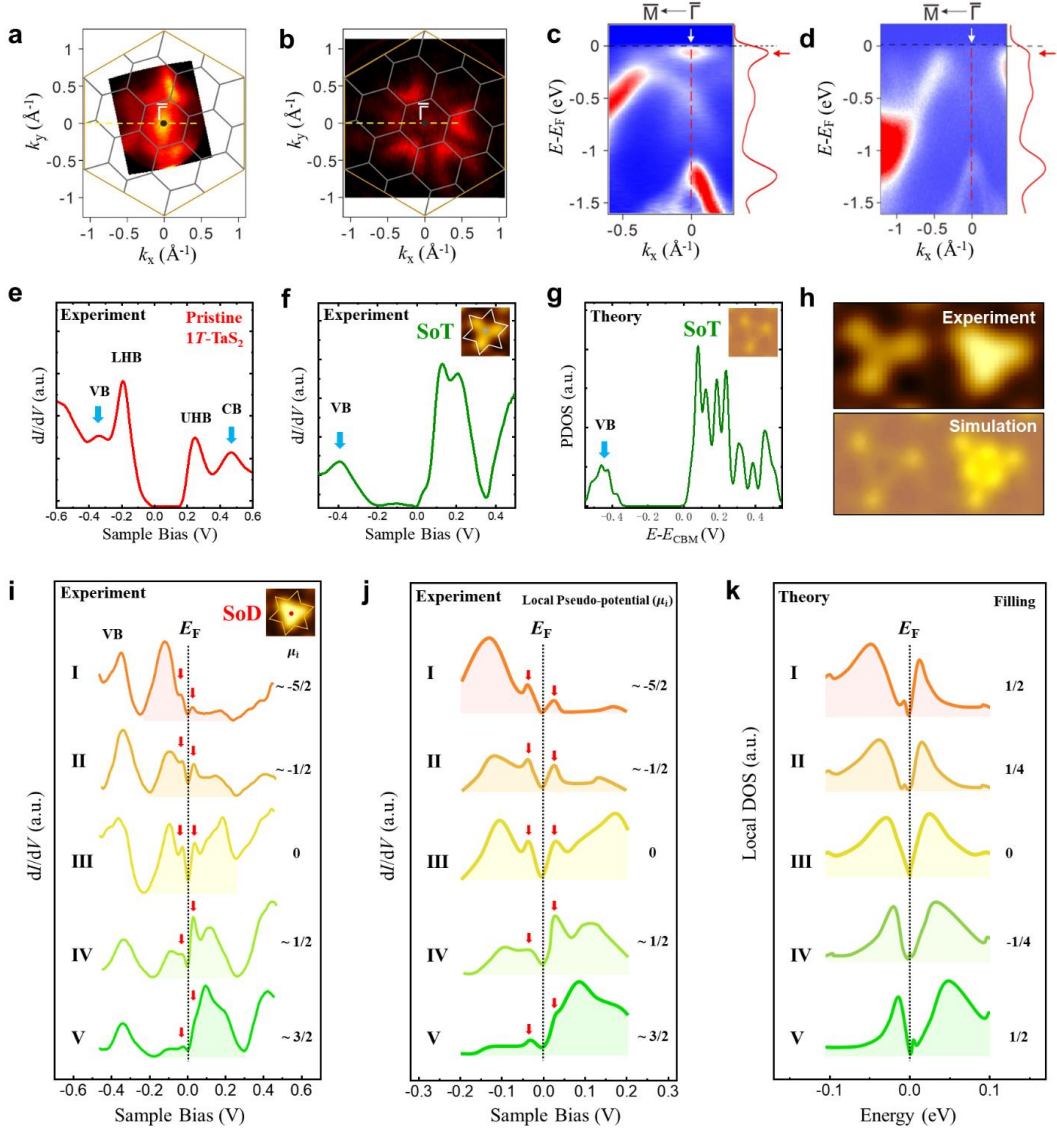

**Figure 2. Electronic structures of hole-doped 1$T$-TaS$_2$.** (a,b) The density of state (DOS) of flat-band at the Fermi level of the medium-hole-doped and high-hole-doped 1$T$-TaS$_2$. (c,d) The energy-dependent band spectra along $\overline{\Gamma} - \overline{M}$ direction of the medium-hole-doped and high-hole-doped 1$T$-TaS$_2$. The energy distribution curves (EDCs) taken at $\overline{\Gamma}$ point is shown by red curves. (e) The d$I$/d$V$ spectra of the SoDs in pristine 1$T$-TaS$_2$, in which the LHB, UHB, VB and CB represent lower Hubbard band, upper Hubbard band, valence band and conduction band respectively. (f,g) The typical d$I$/d$V$ spectra and theoretical DOS of SoTs in hole-doped 1$T$-TaS$_2$. (h) The comparative experimental and simulated images of SoT and SoD. (i) Potential-dependent d$I$/d$V$ spectra of the SoDs in hole-doped 1$T$-TaS$_2$. Local potentials are roughly estimated by the form: $\mu_i \approx \frac{|A_{left}| - |A_{right}|}{2}$. (j) Zoomed-in d$I$/d$V$ spectra of (i), with the two peak (~$\pm$30 mV) marked by red arrows. (k) The dynamical mean field theory (DMFT) simulation of local DOS at different potential $\mu_i$. [46].

The electronic structures of particle-like correlated electrons in hole-doped 1$T$-TaS$_2$ are further investigated by the ***k***-space ARPES and ***r***-space STS spectra measurements. Figure 2a and b depict the measured density of states (DOS) of flat-bands at the Fermi level in the medium-hole-doped and high-hole-doped 1$T$-TaS$_2$, respectively. The clear and faint Fermi surface intensities are observed at the $\bar{\Gamma}$ point, respectively, consistent with their partial- and tiny-filled flat-bands of these correlated electrons. Figure 2c and d display the energy-dependent band spectra along $\bar{\Gamma} - \bar{M}$ directions for the medium-hole-doped and high-hole-doped 1$T$-TaS$_2$, respectively. Clear and faint peaks near the Fermi level (indicated by red arrows) is also respectively observed in the energy distribution curves (EDCs) around the $\bar{\Gamma}$ point. It is noted that no clear Hubbard band splitting can be resolved in these hole-doped 1$T$-TaS$_2$ due to the reduced correlation effect of these flat-band electrons, different from the pristine 1$T$-TaS$_2$. The upward-shifting CDW-occupied bands can also be clearly resolved, consistent with the elevated potentials due to the hole-doping effect of element-substitution.

In pristine 1$T$-TaS$_2$, the STS spectra of electron-filled SoDs display a prominent Hubbard gap due to the strong $U$, as shown in Fig. 2e. In contrast, the electron-filled SoDs and electron-empty SoTs demonstrate two starkly distinct STS spectra in hole-doped 1$T$-TaS$_2$. The representative STS spectra of SoTs are displayed in Fig. 2f, with the spectral weights of occupied state suppressed and transferred to the empty state, aligning with the theoretical DOS of SoT (Fig. 2g and Supplementary Fig. 4). Figure 2h illustrates comparative the experimental and simulated images of SoT and SoD, coinciding with the spectra results. Different from the identical electron-empty SoTs, the electron-filled SoDs demonstrate clear position-specific spectra (spectra I-V), as shown in Fig. 2i, due to the different local potential variations caused by the varying surrounding SoDs conditions.

All SoDs exhibit high DOS near the Fermi level, forming a V-shaped pseudogap without the pronounced Hubbard gap due to the small $U$, distinct from the pristine 1$T$-TaS$_2$. The zoomed-in STS spectra further identify the positions of the two peaks (~±30 mV) with a separation of about 60 mV (Fig. 2j), which are consistent across different SoDs. Meanwhile, it is also noted that the position of the VB of spectra I-V remains essentially unchanged, while the spectral weight near the Fermi level gradually shifts from the occupied to the empty states. Similar spectral evolution has been observed in doped Mott insulating phases in cuprates [47] and in twisted graphene bilayers [48], attributed to doping-induced and gating-induced local potential variations, respectively. Furthermore, the local DOS with different potential variations calculated by dynamical mean field theory (DMFT) in magic-angle twisted bilayer graphene [46], as shown in Fig. 2k, is also in accordance with our spectra results. These results indicate that the particle-like electrons in hole-doped 1$T$-TaS$_2$ show reduced electronic correlation (small $U$) and potential-dependent electronic structure.

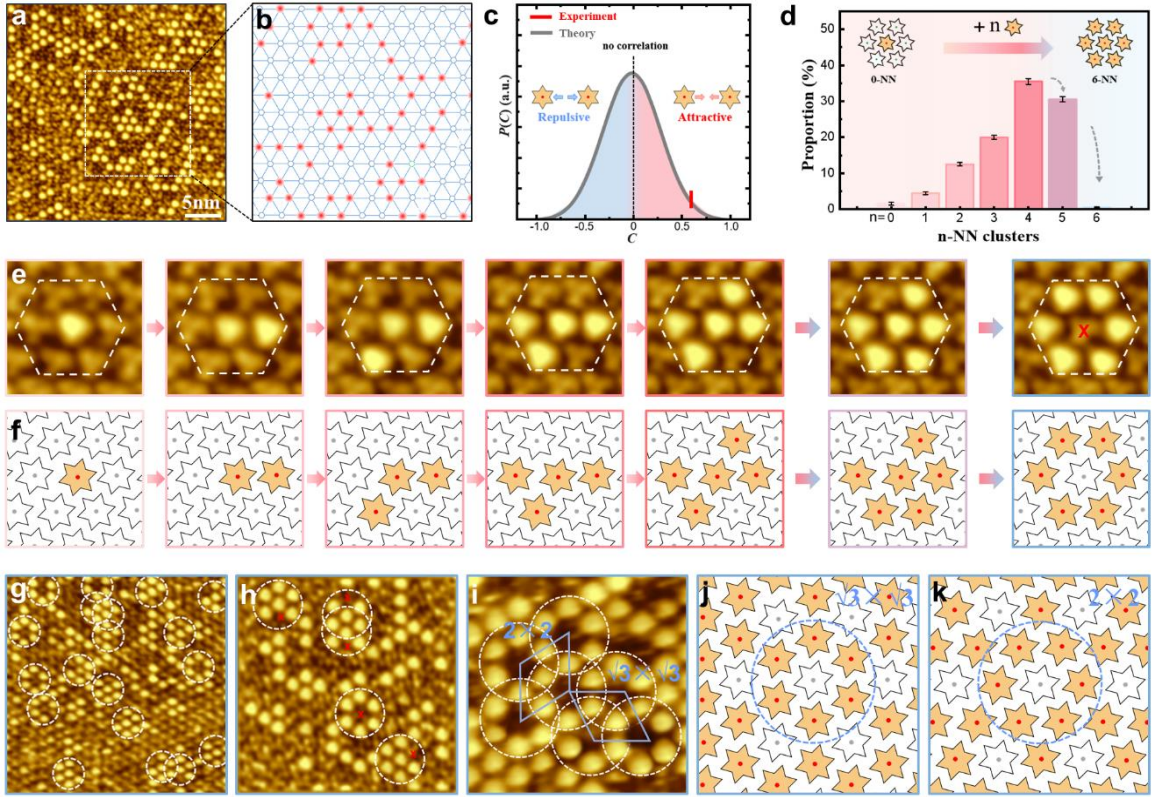

**Figure 3. Spatial distributions of particle-like correlated electrons in hole-doped 1T-TaS$_2$.** (a) Large-scale STM image of distributed correlated electrons (electron-filled SoDs). (b) Schematic distribution of electron-filled (SoD, red circle) and electron-empty (SoT, grey circle) in dotted box of (a). (c) The nearest-neighbor spatial correlator C calculated from theory [49] and experiment. (d) Histogram counting of electron clustering configurations (n-NN clusters) from 0-NN to 6-NN, with NN representing nearest-neighboring electrons. (e,f) High-resolution STM images (e) and their corresponding schematic (f) of electron clustering configurations, indicating the existence of nearest-neighbor attractive interactions (0-NN to 4-NN) and many-body repulsive interaction (6-NN). (g,h) Typical STM images of anti-7 magic number configuration highlighted by the white circles. (i-k) STM image (i) and schematic (j,k) of the short-range order (√3×√3/ 2×2 super-superlattice) of anti-7 configurations. Scanning parameter: (a,e,h,i) *V*=-1.0 V, *I*=-100 pA; (g) *V*=-1.2 V, *I*=-100 pA.

The correlated electron systems in hole-doped Mott insulator have been theoretically described and discussed by the extended Hubbard model, which is written as the form:

$$H = \sum_{i,j} t_{i,j} \hat{C}_i^\dagger \hat{C}_j + U \sum_i n_{i,\uparrow} n_{i,\downarrow} + \sum_i \mu_i \hat{C}_i^\dagger \hat{C}_i + \sum_{i,j} V_{i,j} \hat{n}_i \hat{n}_j + \sum_{i_1,i_2,\cdots i_7} W_{i_1,i_2\cdots,i_7} \hat{n}_{i_1} \cdots \hat{n}_{i_7}$$

, where $V_{i,j}$ parametrizes the nearest neighbor interactions, $W_{i_1,i_2\cdots,i_7}$ denotes multi-body interaction and $\hat{n}_i = \hat{C}_i^\dagger \hat{C}_i$ is the number operator in site i. The Coulomb repulsion *U* and the potential $\mu_i$ of the particle-like electrons in hole-doped 1T-TaS$_2$ have been discussed in Fig. 2, and now we focus on the nearest-neighbor interaction $V_{i,j}$ and the multi-body interaction $W_{i_1,i_2\cdots,i_7}$ in Fig. 3. These two correlated interactions are crucial for understanding the complex spatial distributions of particle-like electrons, but they have been less studied.

To elucidate the correlated interactions between the particle-like electrons, the relative spatial distributions of electrons are statistically analyzed. Figure 3a shows a typical STM image of distributed correlated electrons (electron-filled SoDs), which are schematically illustrated by the red circles in the triangular CDW superlattice in Fig. 3b. It is clear that these electrons are not randomly distributed, but in cluster or short-chain configurations (Supplementary Figs. S5 and S6), indicating the attractive interactions between the electrons. Their relative spatial distribution can be semi-quantitatively described by the spatial correlators [49]. The nearest-neighbor spatial correlator can be defined as $C=<n_i n_j>_{ij\in NN} - <n_i><n_j>$, where $n_i$ =1, 0 for an electron-filled, electron-empty site i, respectively, ij $\in$ NN denotes nearest-neighbor pairs in the triangular superlattice and $<>$ denotes the statistical average. For an infinitely large system with a random distribution of electrons and without any correlations, the nearest-neighbor correlator C would be identically 0. The positive correlator C calculated from our STM data, as shown in Fig. 3c, quantitatively indicates the presence of the prominent attractive nearest-neighbor interactions between electrons. Notably, the observed nearest-neighbor attractive interaction $V_1$ is larger than previously reported values in the 1T/1H-TaS$_2$ heterostructures [49], which may stem from the Coulomb screening effect of the 1*H*-TaS$_2$ layer on the electrons of the 1*T*-TaS$_2$ layer.

An unexpected many-body interaction term of these correlated electrons is further discovered by a contrastive statistical analysis of their clustering configurations, as shown in Fig. 3d-f. These electron clustering configurations, named as n-NN clusters, are defined as the target electron with n nearest-neighboring (NN) electrons, and further counted from our STM data. Histogram analysis of their proportions is supplied in Fig. 3d, in which the proportions of n-NN gradually increase from 0-NN to 4-NN, consistent with the attractive NN interactions. However, the proportions of 5-NN gently decreased, and almost no 6-NN electron clusters are found, indicating the existence of a noteworthy repulsive many-body interaction term for 6-NN cluster. The non-existence of 6-NN cluster, named anti-7 magic number configuration, is directly displayed by the marked white circles in the STM images of Fig. 3g and h. This specific 6-NN repulsive interaction term can also lead to the short-range order of electron-vacancy (anti-7 configuration) in the √3×√3 or 2×2 super-superlattice of CDW superlattice (Fig. 3i-k and Supplementary Fig. 7). The similar super-modulations have also been reported in the correlated electronic states of monolayer 1$T$-TaS$_2$ (and 1$T$-NbSe$_2$, 1$T$-TaSe$_2$), as shown in Supplementary Fig. 8, implying their possible consistent underlying mechanism [50-54]. As a unique many-body interaction term experimentally discovered in real-space, more theoretical work is needed to understand its atomic mechanism and predict its effect on the emergent quantum states in the extended Hubbard model framework.

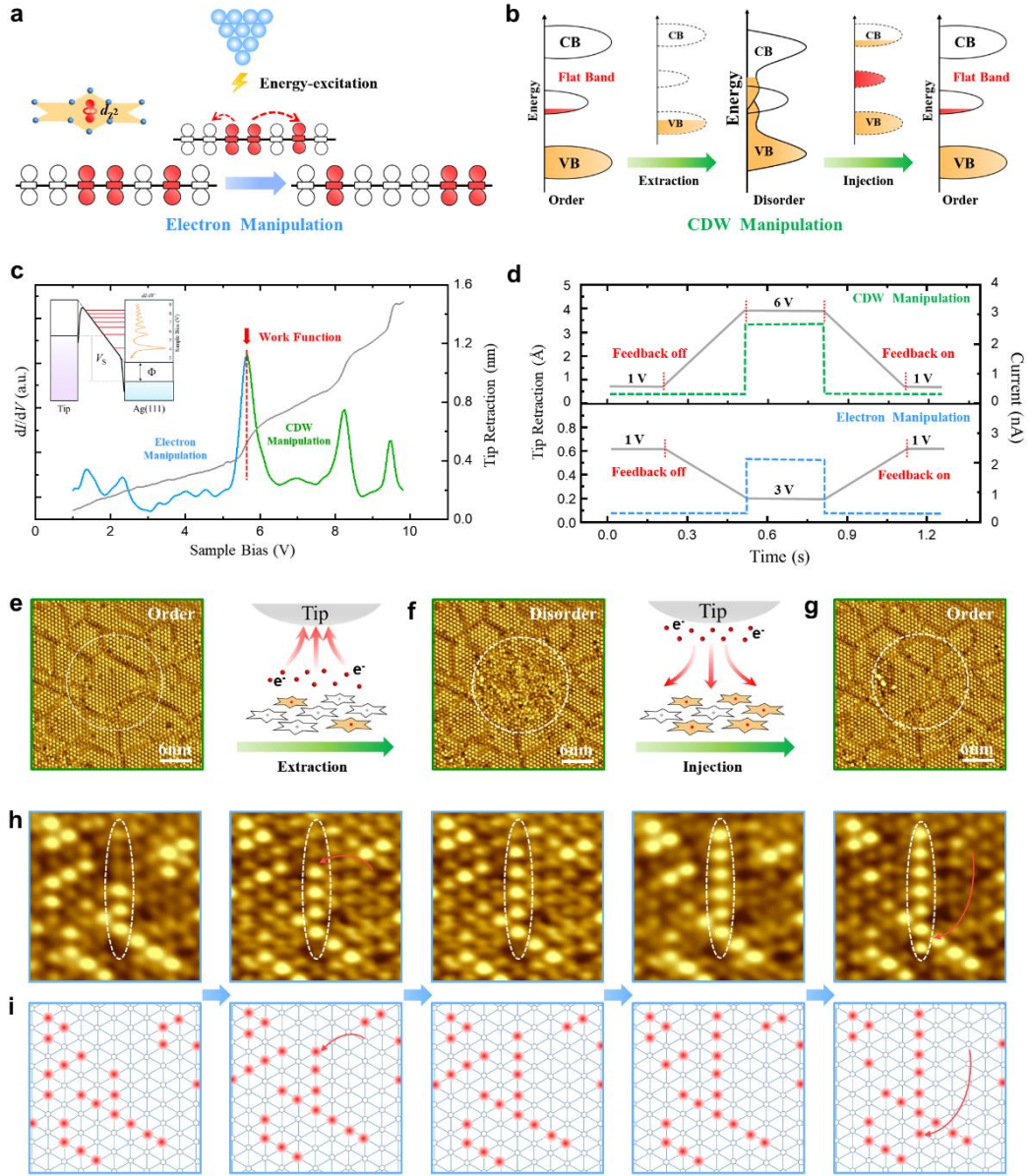

**Figure 4. Local manipulations of particle-like correlated electrons in hole-doped 1*T*-TaS$_2$.** (a) Schematic of energy-excitation-based electron manipulation mechanism. (b) Schematic illustrations of the electron-filling-based CDW disordering-reordering manipulation. (c) Field emission resonance (FER) spectra of the hole-doped 1*T*-TaS$_2$. The blue and green lines indicate the voltage ranges for electron manipulation and CDW manipulation, respectively. The inset gives the FER mechanism and spectra of Ag (111) surface. (d) Manipulation processes of particle-like correlated electrons and CDW. (e-g) STM images of local CDW ordering (e), disordering (f) and reordering (g) manipulations by the large positive and negative tip bias pulse (>5.5 V). The cartoons illustrate the transient electron-extraction and electron-injection process. (h,i) A sequence of STM images and their schematics of local particle-like electron manipulations by the small positive and negative tip bias pulse (<5.5 V). Scanning parameter: (e,f,g) *V*= -0.5 V, *I*= -100 pA; (h) *V*= -1 V, *I*= -100 pA.

Manipulating single electrons at the atomic scale is vital for understanding the correlated system dominated by particle-like electrons. Figure 4a illustrates a schematic of the proposed manipulation mechanism of particle-like correlated electrons in hole-doped $1T$-$TaS_2$, in which the electrons are energetically excited by the tunneling electrons to jump into the nearby vacancy sites (electron-empty SoTs). In addition, the electron-filled CDW bands can be temporarily modulated by extracting electrons from the sample (or injecting electrons into the sample) to achieve the CDW ordered-disordered transition, as shown in Fig. 4b. The global CDW disordering manipulation has been realized by the high-energy laser pulse [55], while the localized disordering-reordering CDW manipulations here can be realized by the controlled electron extraction-injection process via the positive-negative bias pulses of STM tip.

We demonstrated the manipulation of particle-like correlated electrons and CDW with the STM-based field emission resonance (FER) spectroscopy measurement. The work function (~5.5 eV) of hole-doped $1T$-$TaS_2$ is determined by the 1st resonance peak of FER spectroscopy in Fig. 4c. It is found that applying large positive and negative tip bias pulse (>5.5 V, larger than its work function) with the feedback loop off can realize local CDW disordering and reordering manipulations rather than electron manipulation, as illustrated in Fig. 4d-g, and Supplementary Figs 9 and 10. To achieve the manipulation of particle-like correlated electrons, the small STM bias pulses (<5.5 V) is applied with the feedback loop off, detecting the tunneling current jump from the low current state (~100 PA) to the high current state (~2 nA), as shown in Fig. 4d. Figure 4h and i display the process of particle-like electron manipulations for the first time. Interestingly, both small positive and negative bias pulses (large tunneling current) can enable the electron hopping manipulation, confirming the proposed energy-excitation mechanism.

Furthermore, the discovered electron correlations in Fig. 3 also play important roles in the electron hopping process, making them more inclined to form electron chains or clusters (Supplementary Figs. 11 and 12), consistent with the nearest-neighbor attractive interactions between electrons. During the energy-excited electron manipulations, the 6-NN electron cluster is occasionally observed and quickly relaxed into the anti-7 electron cluster, consistent with their repulsive 6-NN interactions (Supplementary Fig. 13). Different from the established atom or molecule manipulation techniques on surfaces, pioneered in the 1990s [43, 44], the STM-based particle-like electron manipulation is still in its infancy here. More experimental and theoretical work are needed in this field to realize the controllable and precise electron manipulations. Precise electronic manipulation is conducive to generating novel electron patterns, constructing complex artificial electronic states, and opening the way for exploring entirely new quantum states of matter and functional devices

**Discussion**

The behavior and interactions of particle-like electrons in correlated materials are crucial for understanding the fundamental nature of diverse exotic quantum states. The particle-like electrons are successfully realized and directly visualized in $1T$-TaS$_2$ via precise hole doping. The correlated interactions among these particle-like electrons are further described and discussed within the extended Hubbard model framework. ARPES, STS and theoretical calculations demonstrate the small Coulomb repulsive interaction $U$ and the local potential $\mu_i$ of single particle-like electron. The nearest-neighbor attractive interaction $V_1$ and the many-body repulsive interaction $W$ are quantitatively identified through specific electron clustering configurations. These complex interactions play a key role in the theoretical construction and prediction of novel quantum states in correlated electron systems.

Furthermore, we achieve local artificial manipulation of particle-like electrons at the atomic scale in real space, marking a transition from observation to active manipulation of electrons. This single-electron manipulation capability differs from traditional atomic/molecular manipulation and establishes a new paradigm to modulate correlated electronic states. Although current manipulation remains constrained by many-body correlation effects, optimizing the tuning parameters is expected to enable precise single-electron manipulation in the future. This will provide an effective route to construct artificial electron lattices, explore emergent quantum states, and develop quantum functional devices.

In summary, we report the spatial distribution, electronic structure, interactions, and preliminary manipulation of particle-like electrons in doped Mott insulators. The doping-dependent average electronic structure and the potential-dependent local electronic structures of single particle-like electrons are investigated by ARPES and STS. The spatial distribution of electrons demonstrates complex electron clustering configurations including small cluster/short chain and anti-7 magic number configurations, quantitatively reflecting the nearest-neighbor attractive and many-body repulsive interactions between correlated electrons. Remarkably, the tentative manipulation of the particle-like electrons and CDW are further realized by energy-excitation and charge-injection mechanisms. Our work not only provides significant insights into particle-like correlated electrons, but also establishes a versatile platform for actively designing and controlling quantum states at the atomic scale.

## Acknowledgments

This project was supported by the National Key R&D Program of China (MOST) (Grant No. 2023YFA1406500), the National Natural Science Foundation of China (NSFC) (No. 92477128, 92477205, 12374200, 11604063, 11974422, 12104504), the Strategic Priority Research Program (Chinese Academy of Sciences, CAS) (No. XDB30000000), the Fundamental Research Funds for the Central Universities and the Research Funds of Renmin University of China (No. 21XNLG27), and the Hainan Provincial Natural Science Foundation of China (No. 525QN341). Y.Y. Geng was supported by the Outstanding Innovative Talents Cultivation Funded Programs 2023 of Renmin University of China.

## Author contributions

H.G., W.J., S.W., W.Z. and Z.C. conceived the research project. Y.G., H.D., and Z.C. performed the STM experiments and analysis of STM data. R.W. and W.J. performed the DFT calculations and analyses. Z.W. and S.W. performed ARPES experiments and analyses. J.G., S.M., Y.L., F.P., R.X. and L.H. helped in the experiments.

## Competing Interests

The authors declare no competing financial interests.

## Data Availability

The authors declare that the data supporting the findings of this study are available within the article and its Supplementary Information.

## Materials and Methods

### Single crystal growth of Hole-doped 1$T$-TaS$_2$.

The high-quality 1$T$-TaS$_2$ and hole-doped 1$T$-TaS$_2$ crystals were grown by chemical vapor transport (CVT) method with iodine as a transport agent. Large hole-doped 1$T$-TaS$_2$ flakes with size up to 10 mm were collected for further characterization and measurement. X-ray diffraction (XRD), scanning electron microscopy (SEM), X-ray energy dispersive spectroscopy (EDS) were employed to determine the crystal structure, morphology, and composition of as-prepared samples.

### STM and AFM measurements

High-quality 1$T$-TaS$_2$ and hole-doped 1$T$-TaS$_2$ crystals were cleaved at room temperature in ultrahigh vacuum at a base pressure of $2\times10^{-10}$ Torr, and directly transferred to the cryogen-free variable-temperature STM system (PanScan Freedom, RHK). Chemically etched W tips were used for STM measurement in constant current mode. The tips were calibrated on a clean Ag(111) surface. The STS spectra were measured in constant-height mode using standard lock-in techniques. Gwyddion was used for STM data analysis. The nc-AFM measurements were performed in a commercial LT-STM system (LT-STM/AFM, CreaTec) equipped with an STM/qPlus sensor at 4.5K. The nc-AFM images were recorded by measuring the frequency shift of the qPlus resonator (sensor frequency $f_0$=30kHz, Q= 25000) in constant-height mode with an oscillation amplitude of 180 pm.

### ARPES measurements.

ARPES measurements were performed at Renmin University of China equipped with a Scienta DA30 analyzer, with photon energy of 10.05 eV, the BL03U beamline of Shanghai Synchrotron Radiation Facility (SSRF), and the BL13U beamline of National Synchrotron Radiation Laboratory (NSRL). The energy and angular resolution were set to 10 meV and 0.02 Å$^{-1}$, respectively. Clean surfaces for ARPES measurements were obtained by *in situ* cleaving the samples. Photoemission spectra presented in this work were recorded $T$= 15 K using the photo energy $hv$=23eV and 70eV, in a working ultrahigh vacuum better than $7\times10^{-11}$ Torr.

### Density functional theory calculations and simulations.

Our DFT calculations were performed using the generalized gradient approximation for the exchange-correlation potential, the projector augmented wave method and a plane-wave basis set as implemented in the Vienna ab-initio simulation package (VASP). Dispersion correction was implemented using the DFT-D3 method, with the PBE functional for the exchange potential. The effective $U$ value of the on-site Coulomb interaction of the Ta d orbitals is 2.3 eV. All our simulation cells contain a 20 Å vacuum to prevent interlayer coupling between different layers. The plane-wave cutoff was set to 450 eV. A k-mesh of 5×5×1 was adopted to sample the first Brillouin zone of the cell. Convergence is reached if the consecutive energy difference is within 10-5 eV for electronic iterations and the forces are less than 0.01 eV Å.